\documentclass[10pt,letterpaper,nofonttune]{IEEEtran}
\usepackage{eso-pic}
\usepackage{graphicx}
\usepackage{tikz}
\usepackage{pgf}
\usepackage{lscape}
\usepackage{amssymb}
\usepackage{fancyvrb}
\usepackage{amsfonts}
\usepackage{verbatim}
\usepackage{xfrac}
\usepackage{algorithmic}
\usepackage{url}
\usepackage{amsmath}
\usepackage{multirow}
\usepackage{listings}
\usepackage{xcolor}
\usepackage{algorithm}
\usepackage{bm}
\usepackage{proof}
\usepackage{epstopdf}
\usepackage{subfig}
\usetikzlibrary{arrows,automata}
\usetikzlibrary{shapes,snakes}

\makeatletter
\newcommand*\titleheader[1]{\gdef\@titleheader{#1}}
\AtBeginDocument{%
  \let\st@red@title\@title
  \def\@title{%
    \bgroup\normalfont\large\centering\@titleheader\par\egroup
    \vskip1.5em\st@red@title}
}
\makeatother

\usepackage[numbers,sort&compress]{natbib}

\begin{document}

\title{\ \ \ \ \ \ \ \ \  \textit{Big Data Goes Small:}\newline Real-Time Spectrum-Driven Embedded Wireless Networking Through Deep Learning in the RF Loop}
\titleheader{Accepted to IEEE INFOCOM 2019}

\author{\IEEEauthorblockN{Francesco Restuccia and Tommaso Melodia}\\
\IEEEauthorblockA{Department of Electrical and Computer Engineering\\
Northeastern University\\
Boston, MA 02115 USA\\
Email: \{frestuc, melodia\}@northeastern.edu}
\thanks{This paper has been accepted for publication in IEEE INFOCOM 2019. This is a preprint version of the accepted paper. Copyright (c) 2013 IEEE. Personal use of this material is permitted. However, permission to use this material for any other purposes must be obtained from the IEEE by sending a request to pubs-permissions@ieee.org.}
}

\maketitle

\begin{abstract}
The explosion of 5G networks and the Internet of Things will result in an exceptionally crowded RF environment, where techniques such as spectrum sharing and dynamic spectrum access will become essential components of the wireless communication process. In this vision, wireless devices must be able to (i) learn to autonomously extract knowledge from the spectrum on-the-fly; and (ii) react in real time to the inferred spectrum knowledge by appropriately changing communication parameters, including frequency band, symbol modulation, coding rate, among others. Traditional CPU-based machine learning suffers from high latency, and requires application-specific and computationally-intensive feature extraction/selection algorithms. Conversely, deep learning allows the analysis of massive amount of unprocessed spectrum data without ad-hoc feature extraction. So far, deep learning has been used for offline wireless spectrum analysis only. Therefore, additional research is needed to design systems that bring deep learning algorithms directly on the device's hardware and tightly intertwined with the RF components to enable real-time spectrum-driven decision-making at the physical layer. In this paper, we present \emph{RFLearn}, the first system enabling spectrum knowledge extraction from unprocessed I/Q samples by deep learning directly in the RF loop. \emph{RFLearn} provides (i) a complete hardware/software architecture where the CPU, radio transceiver and learning/actuation circuits are tightly connected for maximum performance; and (ii) a learning circuit design framework where the latency vs. hardware resource consumption trade-off can explored. We implement and evaluate the performance of  \emph{RFLearn} on custom software-defined radio built on a system-on-chip (SoC) ZYNQ-7000 device mounting AD9361 radio transceivers and VERT2450 antennas. We showcase the capabilities of \emph{RFLearn} by applying it to solving the fundamental problems of modulation and OFDM parameter recognition. Experimental results reveal that \emph{RFLearn} decreases latency and power by about 17x and 15x with respect to a software-based solution, with a comparatively low hardware resource consumption.
\end{abstract}

\IEEEpeerreviewmaketitle


\section{Introduction and Motivation}\label{sec:intro}

Today's spectrum environment is exceptionally crowded. According to the latest Ericsson's mobility report, there are now 5.2 billion mobile broadband subscriptions worldwide, generating more that 130 exabytes per month of wireless traffic \cite{EricssonMobility2018}. Moreover, it is expected that by 2020, over 50 billion devices will be absorbed into the Internet, generating a global network of ``things'' of dimensions never seen before \cite{CiscoEstimates}. 

Given that only few radio spectrum bands are available to wireless carriers \cite{SpectrumCrunch}, technologies such as radio-frequency (RF) spectrum sharing through beamforming \cite{shokri2016spectrum,vazquez2018hybrid,lv2018cognitive},  dynamic spectrum access \cite{jin2018specguard,chiwewe2017fast,FederatedWireless,agarwal2016edsa} and anti-jamming technologies \cite{zhang2017framework,huang2017anti,chang2017jamming} will become essential in the near future. The first key challenge in enabling these systems is how to \textit{effectively} and \textit{efficiently} extract \textit{meaningful}  and \textit{actionable} knowledge out of the tens of millions of in-phase/quadrature (I/Q) samples received every second by wireless devices. To give an example, to monitor a single $\mathrm{20~MHz}$ WiFi channel, we need to process at least 40 million I/Q samples/s at Nyquist sampling rate. This generates a stream rate of about $\mathrm{2.56~Gbit/s}$, if samples are each stored in a 4-byte word. The second core challenge is that the RF channel is significantly time-varying (\textit{i.e.}, in the order of milliseconds), which imposes strict timing constraints on the \textit{validity} of the extracted RF knowledge. If (for example) the RF channel changes every 10ms, a knowledge extraction algorithm must run with latency (much) less than 10 ms to both (i) offer an accurate RF prediction and (ii) drive an appropriate physical-layer response to the inferred spectrum knowledge; for example, change in modulation/coding/beamforming vectors due to adverse channel conditions, local oscillator (LO) frequency due to spectrum reuse, and so forth. 

To address the knowledge extraction challenge, \textit{deep learning} \cite{lecun2015deep} has been widely recognized as the technology of choice for solving classification problems for which no well-defined mathematical model exists. Deep learning goes beyond traditional low-dimensional machine learning techniques by enabling the analysis of unprocessed I/Q samples without the need of application-specific and computational-expensive feature extraction and selection algorithms \cite{OShea-ieeejstsp2018}. Another core advantage is that deep learning architectures are application-insensitive, meaning that the same architecture can be retrained for different learning problems. For this reason, the Defense Advanced Research Projects Agency (DARPA), among other agencies in the United States, has recently launched the novel radio-frequency machine learning systems (RFMLS) research program \cite{DARPA}, where the main objective is to fingerprint wireless devices by learning from RF data, rather than designing ad hoc systems hand-engineered by experts.

The application of learning techniques to the RF domain presents several major hurdles that are substantially absent in traditional learning domains. Indeed, deep learning has been traditionally used in static contexts (\textit{e.g.}, image and language classification \cite{krizhevsky2012imagenet,hinton2012deep}), where the model latency is usually not a concern.  \textit{Another fundamental issue absent in traditional deep learning is the need to satisfy strict constraints on resource consumption.} Indeed, models with high number of neurons/layers/parameters will necessarily require additional hardware and energy consumption, which are clearly scarce resources in embedded systems. 

Although prior work has investigated the opportunity of using learning \cite{Pawar-ieeetifs2011,Shi-ieeetcomm2012,Xu-ieeetvt2010} and deep learning \cite{Kulin-ieeeaccess2018,OShea-ieeejstsp2018,Karra-ieeedyspan2017,o2017introduction} techniques for RF spectrum analysis, we are not aware of practical demonstrations of real-time deep learning in the RF loop for spectrum-driven wireless networking on embedded systems. This is not without a reason. The core issue in enabling real-time deep spectrum learning on embedded devices is the existing lack of an embedded software/hardware architectural design where I/Q samples are directly read from the RF front-end and analyzed in real time on the device's hardware without CPU involvement. To further complicate matters, this architecture must also be flexible enough to be reconfigurable through software based on the wireless application's need. Finally, the strict constraints on latency and resource consumption (hardware and energy) imposed by the embedded RF domain necessarily require a design flow where learning performance is also met by energy/latency/hardware efficiency. 

To fill this research gap, this paper makes the following core contributions:\vspace{0.1cm}

$\bullet$ We propose \emph{RFLearn}, the first learning-in-the-RF-loop system where spectrum-driven decisions are enabled through real-time deep learning algorithms implemented directly on the device hardware and operating on unprocessed I/Q samples. \emph{RFLearn} provides (i) a full-fledged  hardware architecture for system-on-chip (SoC) devices binding together CPU, radio  transceiver  and  learning/actuation  circuits for maximum performance (Section \ref{sec:hwarch});  and  (ii)  a novel framework for RF deep learning circuit design that translates the learning model from a software-based implementation to an \emph{RFLearn}-compliant circuit using high-level synthesis (HLS) (Section \ref{sec:dlcore}), where the  constraints on latency, energy, learning, and hardware performance can  be tuned  based  on the application;\vspace{0.1cm} 

$\bullet$ We extensively evaluate \emph{RFLearn} and its design cycle on a custom software radio composed of a  Zynq-7000 SoC mounting AD9361 radio transceivers and VERT2450 antennas (Section \ref{sec:res:modrec}). As a practical case study, we consider the fundamental problem of   modulation and OFDM parameter recognition through deep learning \cite{OShea-ieeejstsp2018}, and train several classifier architectures to address it (Section \ref{sec:res:mod_train}). We experimentally compare the latency and power consumption performance of \emph{RFLearn} with respect to the same model implemented in software (Section \ref{sec:res:hwvssw}). We also apply our design framework to explore the tradeoff between HLS optimization and hardware consumption (Section \ref{sec:res:opt}). Experimental results indicate that \emph{RFlearn} outperforms the software-based system by decreasing latency and power consumption by respectively 17x and 15x, with a relatively low hardware resource consumption. 


\newcommand{\norm}[1]{\left\lVert #1 \right\rVert}

\section{Background Notions on Deep Learning}\label{sec:background}

 We use boldface upper and lower-case letters to denote matrices and column vectors, respectively. For a vector $\mathbf{x}$, $x_i$  denotes the i-th element, $\norm{\mathbf{x}}$ indicates the Euclidean norm, $\mathbf{x}^\intercal$ its transpose, and $\mathbf{x} \cdot \mathbf{y}$ the inner product of $\mathbf{x}$ and $\mathbf{y}$. For a matrix $\mathbf{H}$, $H_{ij}$ will indicate the (i,j)-th element of $\mathbf{H}$. The notation $\mathcal{R}$ and $\mathcal{C}$ will indicate the set of real and complex numbers, respectively. 



Deep neural networks are mostly implemented as multi-layer perceptrons (MLPs). More formally, an MLP with $L$ layers is formally defined as a mapping $f(\mathbf{x}_i; \bm{\theta}) : \mathcal{R}^{i} \rightarrow \mathcal{R}^{o}$ of an input vector $\mathbf{x}_i \in \mathcal{R}^{i}$ to an output vector $\mathbf{x}_l \in \mathcal{R}^{o}$. The mapping happens through $L$ subsequent transformations, as follows:
\begin{equation}
    \mathbf{r}_{j} = f_j(\mathbf{r}_{j-1}, \theta_j) \hspace{1cm} 0 \le j \le L
    \label{eq:dl_layers}
\end{equation}

where $f_j(\mathbf{r}_{j-1}, \theta_j)$ is the mapping carried out by the $j$-th layer. The vector $\bm{\theta} = \{\theta_1, \ldots, \theta_L\}$ defines the whole set of parameters of the MLP. 

A layer is said to be \emph{fully-connected} (FCL) or \emph{dense} if $f_j$ has the form

\begin{equation}
    f_j(\mathbf{r}_{j-1}, \theta_j) = \sigma(\mathbf{W}_j \cdot \mathbf{r}_{j-1} + \mathbf{b}_j)
\end{equation}

where $\sigma$ is an \textit{activation} function, $\mathbf{W}_j$ is the \emph{weight} matrix and $\mathbf{b}_j$ is the \textit{bias} vector. This function introduces a non-linearity in the mapping processing, which allows for ever complex mappings as multiple layers are stacked on top of each other. Examples of activation functions are linear, \textit{i.e.}, $ \sigma(\mathbf{x}) _i = x_i$, rectified linear unit (RLU), \textit{i.e.}, $ \sigma(\mathbf{x})_i = \mbox{max}(0, x_i)$, and so on. Deep neural networks are generally trained using labeled training data, \textit{i.e.}, a set of input-output vector pairs $(\mathbf{x}_{0, i}, \mathbf{x}^*_{L, i}), 1 \le i \le |\mathbf{S}|$, where $\mathbf{x}^*_{L, i}$ is the desired output of the neural network when  $\mathbf{x}_{0, i}$ is used as input.

Convolutional layers (CVLs) \cite{lecun1989generalization}  address the lack of scalability of FCLs by binding adjacent shifts of the same weights together similar to a filter sliding across an input vector. More formally, a CVL consists of a set of $F$ filters $\mathbf{Q}_f \in \mathbb{R}^{h \times w}, 1 \le f \le F$, where $F$ is also called the layer depth. Each filter generates  a \textit{feature map} $\mathbf{Y}^f \in \mathbb{R}^{n' \times m'}$  from an input matrix $\mathbf{X} \in \mathbb{R}^{n \times m}$ according to the following\footnote{For simplicity, \eqref{eq:filters} assumes input and filter dimension equal to 2. This formula can be generalized for tensors having dimension greater than 2.}:

\begin{equation}
    Y^{f}_{i,j} = \sum_{k=0}^{h-1}\sum_{\ell=0}^{w-1} Q^f_{h-k,w-\ell} \cdot X_{1+s\cdot(i-1)-k, 1+s\cdot(j-1)-\ell}
    \label{eq:filters}
\end{equation}

where $s \ge 1$ is an integer parameter called \textit{stride}, $n' = 1 + \left \lfloor{n + h - 2}\right \rfloor$ and $m' = 1 + \left \lfloor{m+b-2}\right \rfloor$. The matrix $\mathbf{X}$ is assumed to be padded with zeros, \textit{i.e.}, $X_{ij} = 0\ \forall i \not\in [1, n],\ j \not\in [1, m]$. The output dimensions can be reduced by either increasing the stride $s$ or by adding a \textit{pooling layer} (POL). The POL computes a single value out of $p \times p$ regions of $\mathbf{Y}$, usually maximum or average value.  For more details on CNNs, the reader may refer to \cite{goodfellow2016deep}.

CNNs are commonly made up of only four layer types: convolutional (CVL), pooling (POL), fully-connected (FCL), and rectified-linear (RLL). The most common CNN architectures stack a number of CVL-RLU layers, (optionally) followed by POL layers, and repeat this pattern until the input has been merged spatially to a small size. At some point, it is common to transition to FCLs, with the last FCL holding the output (i.e., the classification output). In other words, the most common CNN architectures follow the pattern below:

\begin{equation*}
\mbox{IN} \rightarrow \underbrace{[\overbrace{[\mbox{CVL} \rightarrow \mbox{RLL}]}^{N} \rightarrow \overbrace{\mbox{POL}}^{0\ldots P}]}_{1\ldots M} \rightarrow \overbrace{[\mbox{FCL} \rightarrow \mbox{RLL}]}^{1\ldots K} \rightarrow \mbox{FCL}
\end{equation*}
where $N$, $M$ and $K$ need to be chosen according to the specific classification problem. In computer vision applications, the most common parameters used are $0 < N \le 3$, $M \ge 0$, $0 \le K \le 3$ \cite{szegedy2015going,he2016identity,krizhevsky2012imagenet}. However, networks with very high number of $N$ and $K$ have been proposed to achieve better classification accuracy \cite{he2016deep}.

\section{RFLearn Architecture}\label{sec:hwarch}

\begin{figure}[!h]
    \centering
    \includegraphics[scale=0.115]{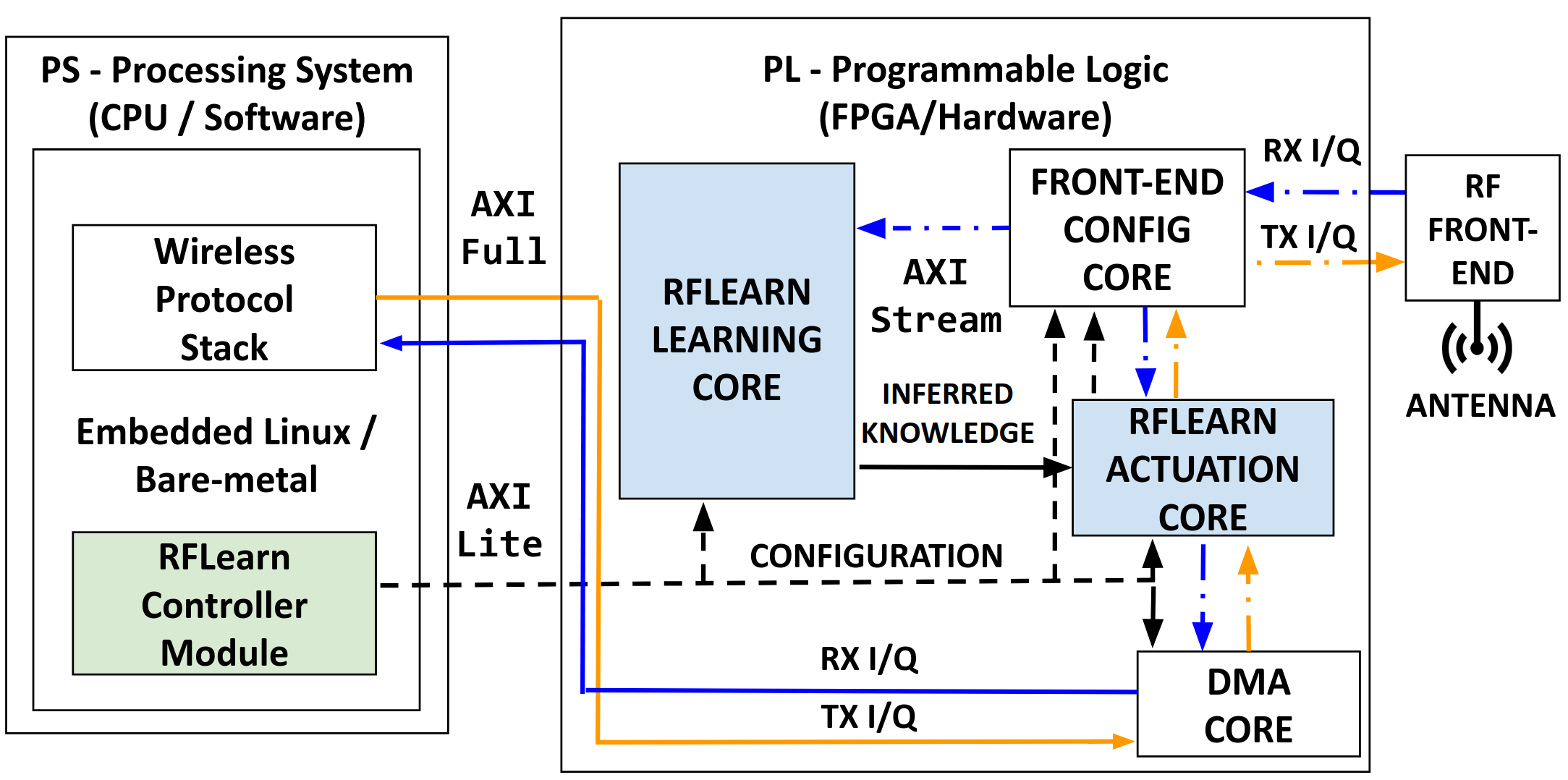}
    \caption{The \emph{RFLearn} Hardware Architecture.\vspace{-0.2cm}}
    \label{fig:rflearn}
    \vspace{-0.3cm}
\end{figure}

Figure \ref{fig:rflearn} depicts a high-level overview of the architecture of the \emph{RFLearn} system.  Together with the RF front-end (hardware) and the wireless network stack (software), \emph{RFLearn} complements a full-fledged  reprogrammable software-defined radio architecture where learning is entirely done in the RF loop without CPU involvement.

We briefly introduce the SoC architecture in Section \ref{sec:soc}, then we describe each and every component of the \emph{RFLearn} system. To ease readability, we have depicted with a blue and yellow color respectively the reception (RX) and transmission (TX) flow of data through the architecture; moreover, configuration data flow has been depicted in black.

\subsection{RFLearn System-on-Chip Computer Architecture}\label{sec:soc}

\emph{RFLearn}'s architectural components entirely reside in the processing system (PS) and the programmable logic (PL) portions of a system-on-chip (SoC), which is an  integrated circuit (also known as ``IC" or ``chip") that integrates all the components of a computer, \textit{i.e.}, central processing unit (CPU), random access memory (RAM), input/output (I/O) ports and secondary storage (\textit{e.g.}, SD card) -- all on a single substrate \cite{Molanes-ieeetie2018}. We refer to SoCs thanks to their low power consumption \cite{SoCLowPower} and because they allow the design and implementation of \textit{customized hardware} on the field-programmable gate array (FPGA) portion of the chip, also called \textit{programmable logic} (PL). Furthermore, SoCs bring unparalleled  flexibility to \emph{RFLearn}, as the PL can be reprogrammed at-will according to the desired learning design. The PL portion of the SoC can be managed by the \textit{processing system} (PS), \textit{i.e.}, the CPU, RAM, and associated buses. 

\emph{RFLearn} uses the \textit{Advanced eXtensible Interface} (AXI) bus specification \cite{XilinxAXI} to exchange data (i) between functional blocks inside the PL; and (ii) between the PS and PL.  We use three AXI sub-specifications in \emph{RFLearn}: \textit{AXI-Lite}, \textit{AXI-Stream} and \textit{AXI-Full}. AXI-Lite is a lightweight, low-speed AXI protocol for register access, and it is used to configure the circuits inside the PL. AXI-Stream is used to transport data between circuits inside the PL. We use AXI-Stream since it provides (i) standard inter-block interfaces; and (ii) rate-insensitive design, since all the AXI-Stream interfaces share the same bus clock, the HLS design tool will handle the handshake between deep learning layers and insert FIFOs for buffering incoming/outgoing samples. AXI-Full is used to enable burst-based data transfer from PL to PS (and \textit{vice versa}). Along with AXI-Full, \emph{RFLearn} uses direct memory access (DMA) to allow PL circuits to read/write data obtained through AXI-Stream to the RAM residing in the PS. The use of DMA is crucial since the CPU would be fully occupied for the entire duration of the read/write operation, and thus unavailable to perform other work. Figure \ref{fig:rflearn} depicts with continuous, dashed, and dot-dashed the AXI-Full, AXI-Lite and AXI-Stream interconnections.


With DMA, the CPU first initiates the transfer, then it does other operations while the transfer is in progress, and it finally receives an interrupt from the DMA controller when the operation is done. This feature is useful when the CPU cannot keep up with the rate of data transfer (which happens very often in the case of RF samples processing). 

\subsection{PS Modules}

In the following, we will refer to as \emph{cores} (or \textit{circuits}) and \emph{modules}, respectively, the \emph{RFLearn} components residing in the PL and PS. 

The main challenge addressed by the PS is to provide modules that will drive and reconfigure the PL cores implementing the learning functionalities provided by \emph{RFLearn}. The PS can run either on top an operating system (such as any embedded Linux distribution), or in "bare-metal" (also called "standalone") mode. In the latter, the only user application running on the CPU is the one specified at compile time. This mode is particularly useful to test the difference in latency between a learning system implemented in the PS (\textit{i.e.}, software) and in the PL (\textit{i.e.}, hardware). 

Through the \emph{RFLearn Controller} module, the PS has full domain over the activities of the cores residing in the PL. Specifically, the \textit{Controller} is tasked to initialize/reconfigure through AXI-Lite (i) the RF front-end core with parameters such as sampling speed, center frequency, finite impulse response (FIR) filter taps, transmission (TX) and reception (RX) local oscillator (LO) frequency, TX/RX RF bandwidth, etc; and (ii) the \emph{RFLearn} learning and actuation cores. The configuration values are stored in registers, so that both the PS and PL cores can access the configuration through memory operations. Moreover, the \emph{Controller} can, at any time, start/stop/check a PL core's operation through registers.

\section{The RFLearn PL Cores}

The main objective of the PL cores is to provide a learning-in-the-RF-loop system where each and every physical-layer operation, included the real-time learning, is done in hardware, with minimum (or absent) involvement of the PS.

The physical-layer data exchange (\textit{i.e.}, I/Q samples) between the PS and PL is handled as follows. The samples flow to/from the PL from/to the PS through the DMA core, which reads/stores the samples from/into the RAM. The wireless protocol stack is tasked with programming the DMA according to its processing rate. However, the DMA can also be configured by the \textit{Controller}, if no wireless protocol stack is present (\textit{i.e.}, the system only processes physical-layer data).

\begin{figure}[!h]
    \centering
    \includegraphics[scale=0.35]{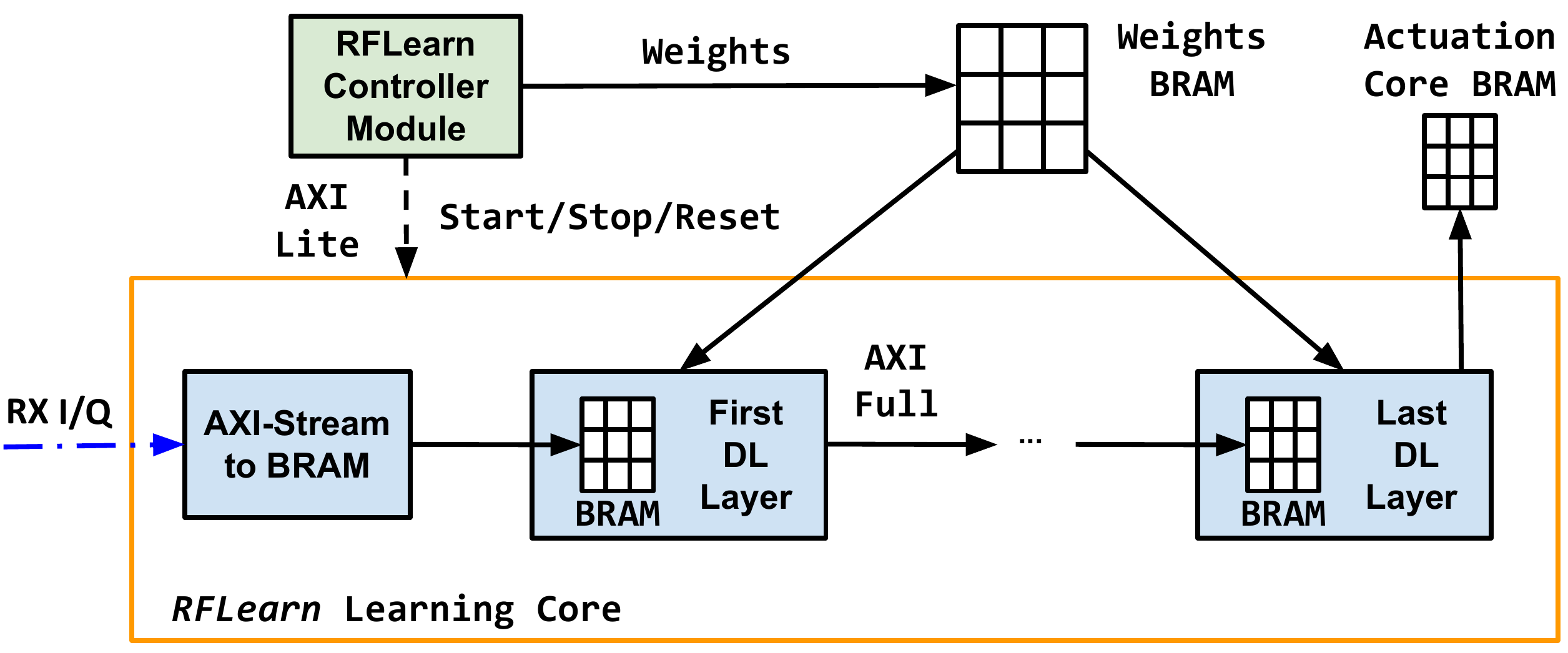}
    \caption{The \emph{RFLearn} Learning Core Architecture.\vspace{-0.2cm}}
    \label{fig:rflearn_learning_core}
\end{figure}

The PL receives/transmits I/Q samples through the \emph{RF Front-end} core, which main operations can be summarized as follows: (i) down/up converts I/Q samples from the carrier frequency (for example, $2.4~\mathrm{GHz}$) to baseband; (ii) applies FIR filtering, DC and I/Q imbalance corrections; and (iii) sends the processed I/Q samples to the \emph{RFLearn Learning} core through AXI-Stream. The I/Q samples received by the \emph{RFLearn Actuation} core go through similar processing before being transmitted over the antenna. As mentioned before, the \emph{RF Front-end} core parameters are set through AXI-Lite, and can be changed both PS-side (\textit{i.e.}, by the \emph{RFLearn} controller) and PL-side by the Actuation core.

The circuit that provides the deep learning capability to the system is the \emph{RFLearn Learning} core, which architecture is detailed in Figure \ref{fig:rflearn_learning_core}. The inputs to this core are (i) a number of unprocessed I/Q samples collected from the radio interface; and (ii) the parameters (\textit{i.e.}, weights, filters, and so on) belonging to each layer (see Equations \ref{eq:dl_layers} and \ref{eq:filters}). Since the core may need to access these quantities in different time instants, both the I/Q samples and the weights are stored in block RAMs (BRAMs), a volatile memory that is implemented entirely in the PL portion of the SoC for maximum speed. Thus, the core necessitates a FIFO that converts the I/Q samples sent through AXI-Stream to a BRAM, so that the core can process the I/Q samples at its own pace. Transactions between the core and the BRAMs are done through AXI-Full. 

Each layer presents the following structure: (i) it receives its input from a BRAM; (ii) it processes the input, according to the type of the layer (\textit{i.e.}, convolutional, fully-connected, rectified linear unit, pooling); (iii) reads the weights from the weights BRAM; (iii) writes the result on the BRAM of the following layer. This architecture presents a number of major advantages: (a) modularity, since layers' computations are independent from each other; (b) scalability, since layers can be added on top of each other without changing the logic of the other layers; (c) reconfigurability, as weights can be changed by the \textit{Controller} at any time \emph{without need to change the hardware structure}. In Section \ref{sec:dlcore}, we will discuss in details how this core is designed and optimized by using HLS.

The \emph{RFLearn Actuation} core has the task to process  the I/Q samples that are received/sent from/to the RF transceiver. Furthermore, the actuation core may (if needed) change the configuration of the RF transceiver itself (\textit{e.g.}, change the FIR filters taps) and the modulation/demodulation logic (\textit{i.e.}, change the physical-layer de-modulation process, increase the coding level, and so on). Since this core's functionality is highly dependent on the given application, we do not propose a specific architecture for it.



\section{The RFLearn DL Core Design Framework}\label{sec:dlcore}

One the fundamental challenges addressed by \emph{RFLearn} is how to transition from a software-based deep learning (DL) implementation (\textit{e.g.}, developed with the Tensorflow \cite{abadi2016tensorflow} engine) to a hardware-based implementation compatible with the \emph{RFLearn} architecture discussed in details in Section \ref{sec:hwarch}. Basic notions of high-level synthesis and the \emph{RFLearn} DL core design are presented in Sections \ref{sec:hls} and \ref{sec:design_steps}, respectively.

\subsection{High-level Synthesis}\label{sec:hls}

\emph{RFLearn} uses high-level synthesis (HLS) for its core designs. HLS is an automated design process that interprets an algorithmic description of a desired behavior (\textit{e.g.}, C/C++) and creates a model written in hardware description language (HDL) that can be executed by the FPGA and implements the desired behavior \cite{winterstein2013high}. 

Designing digital circuits using HLS has several advantages over traditional approaches. First, HLS programming models can implement almost any algorithm written in C/C++. This allows the developer to spend less time on the HDL code and focusing on the algorithmic portion of the design, and at the same time avoid bugs and increase efficiency, since HLS optimizes the circuit according to the system specifications. 

The clock speed of today's FPGAs is several orders of magnitude slower than CPUs (\textit{i.e.}, up to 200-300 MHz in the very best FPGAs). Thus, parallelizing the circuit's operations is crucial. In traditional HDL, transforming the signal processing algorithms to fit FPGA's parallel architecture requires challenging programming efforts. On the other hand, an HLS toolchain can tell how many cycles are needed for a circuit to generate all the outputs for a given input size, given a target parallelization level. This helps \emph{RFLearn} to make the best trade-off between hardware complexity and latency.\vspace{0.1cm}

\emph{Loop Pipelining.}  In high-level languages (such as C/C++) the operations in a loop are executed sequentially and the next iteration of the loop can only begin when the last operation in the current loop iteration is complete. \emph{RFLearn} uses loop pipelining to allows the operations in a loop to be implemented in a concurrent manner.

\begin{figure}[!h]
    \centering
    
\begin{lstlisting}
for (int i=0; i<2;i++) {
    Op_Read;    /* RD */
    Op_Execute; /* EX */
    Op_Write;   /* WR */
}
\end{lstlisting}
    \includegraphics[scale=0.32]{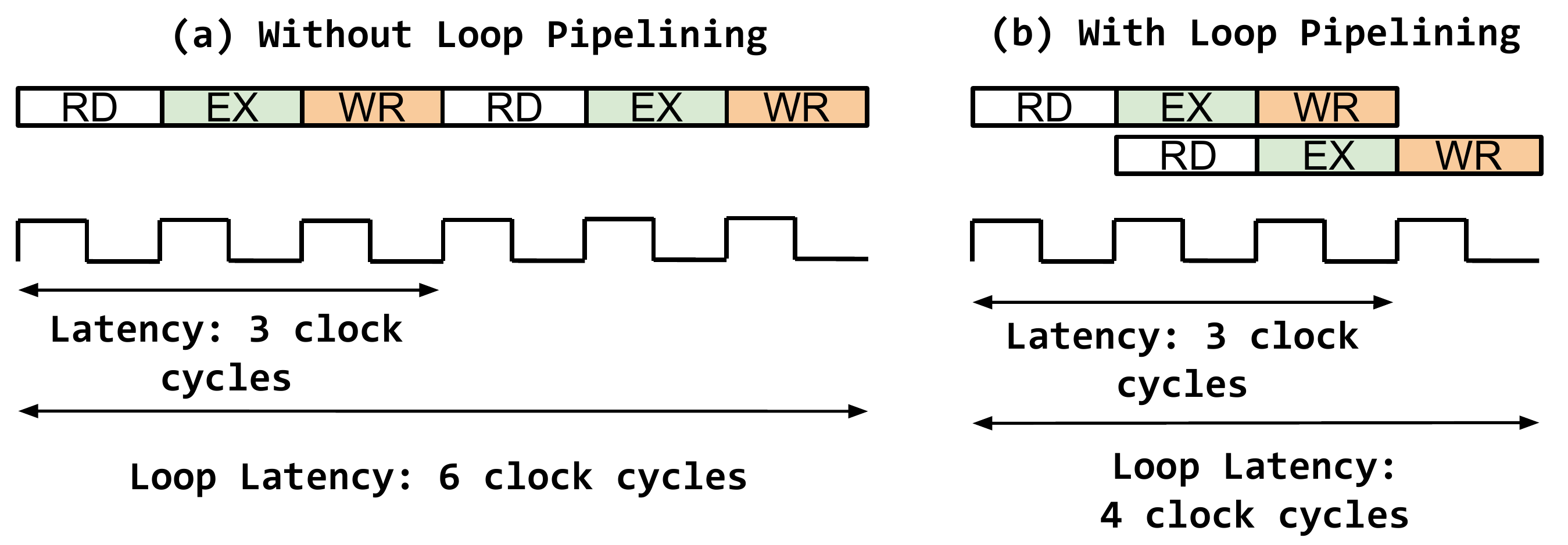}
    \caption{Loop pipelining.\vspace{-0.2cm}}
    \label{fig:pipelining}
\end{figure}

Figure \ref{fig:pipelining} shows an example of loop pipelining, where a simple loop of three operations, \textit{i.e.}, read (RD), execute (EX), and write (WR), is executed twice. For simplicity, we assume that each operation takes one clock cycle to complete. Without loop pipelining, the loop would take 6 clock cycles to complete. Conversely, with loop pipelining, the next RD operation is executed concurrently to the EX operation in the first loop iteration. This brings the total loop latency to 4 clock cycles. If the loop length were to increase to 100, then the latency decrease would be even more evident: 300 versus 103 clock cycles, corresponding to a speedup of about 65\%.

\begin{figure}[!h]
    \centering
    \includegraphics[scale=0.31]{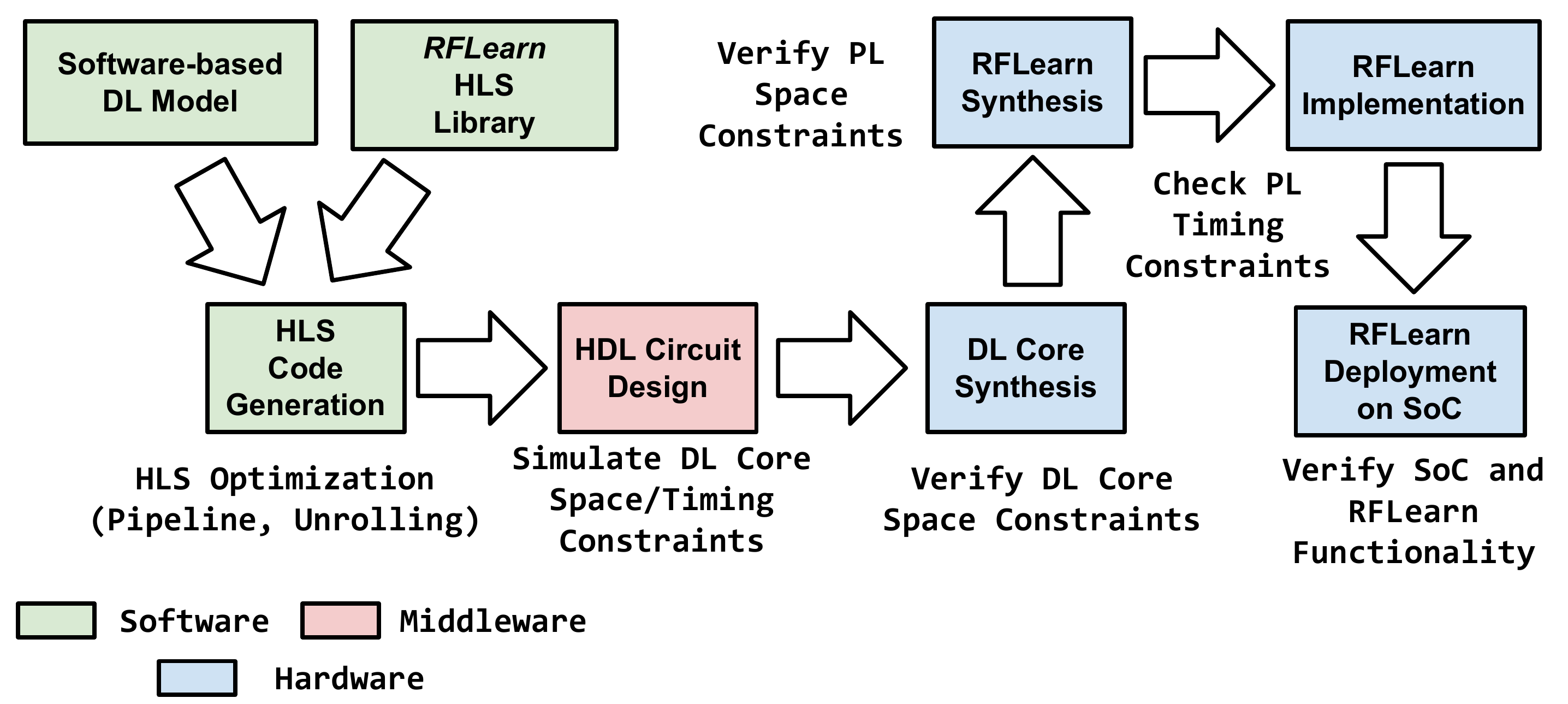}
    \caption{The \emph{RFLearn} DL Core Design Framework.}
    \label{fig:rflearn_frame}
    \vspace{-0.3cm}
\end{figure}

\subsection{Design Steps}\label{sec:design_steps}

The framework presents several design and development steps, which are illustrated in Figure \ref{fig:rflearn_frame}. Steps that involve hardware, middleware (\textit{i.e.}, hardware description logic, or HDL), and software have been depicted with a blue, red, and green shade, respectively. 


The first major step of the framework is to take an existing DL model and convert the model in HLS language, so it can be optimized and later on synthesized in hardware. Another critical challenge is how to make the hardware implementation fully reconfigurable, \textit{i.e.}, the weights of the DL model may need to be changed by the \textit{Controller} according to the specific training. To address these issues, \emph{RFLearn} distinguishes between (i) the DL model architecture, which is the set of layers and hyper-parameters that compose the model itself, as in Equation \eqref{eq:dl_layers}; and (ii) the parameters of each layer, i.e., the neurons' and filters' weights (see Section \ref{sec:background} for details). 

To generate the HLS code describing the software-based DL model, we leverage our own \textit{RFLearn HLS Library}, which provides a set of HLS functions that parse the software-based DL model architecture and generates the HLS design corresponding to the architecture depicted in Figure \ref{fig:rflearn_learning_core}. The \textit{RFLearn HLS Library} currently supports the generation of convolutional (CVL), fully-connected (FCL), rectified linear unit (RLU), and pooling (POL) layers, and operated on fixed-point arithmetic for better latency and hardware resource consumption. The HLS code is subsequently translated to HDL code by an automated tool that takes into account optimization directives such as loop pipelining and loop unrolling. At this stage, the HDL describing the DL core can be simulated to (i) calculate the amount of PL resources consumed by the circuit (\textit{i.e.}, flip-flops, BRAM blocks, etc); and (ii) estimate the circuit latency in terms of clock cycles.

After a compromise between space and latency as dictated by the application has been found, the DC core can be synthesized and integrated with the other PL components of \emph{RFLearn}, and thus total space constraints can be verified. After implementation (\textit{i.e.}, placing/routing), the PL timing constraints can be verified, and finally the whole \emph{RFLearn} system can be deployed o and its functionality tested.

\section{Case Studies: Modulation and OFDM Parameters Recognition}\label{sec:res:modrec}

To evaluate the performance of \emph{RFLearn} on two real-world RF deep learning (DL) problems, we have considered the problem of physical-layer modulation recognition (in short, \textit{ModRec}) and OFDM parameter recognition (in short, \textit{OFDMRec}). We chose these problems since they are fundamental toward  understanding  ongoing wireless transmissions on a given portion of the spectrum.  



\begin{figure}[!h]
    \centering
    \includegraphics[scale=0.9]{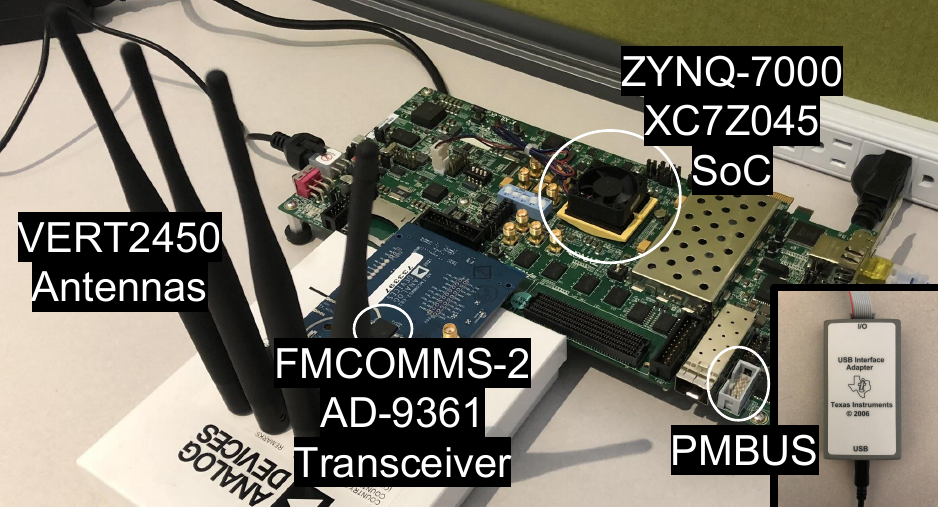}
    \caption{The \emph{RFLearn} Experimental Testbed.}
    \label{fig:zc706}
    \vspace{-0.3cm}
\end{figure}

For our experimental evaluation, we implemented the testbed shown in Figure \ref{fig:zc706} and composed of the following pieces of equipment: (i) a Xilinx Zynq-7000 XC7Z045-2FFG900C system-on-chip (SoC) with two ARM Cortex-A9 MPCore CPUs as processing system (PS) and a Kintex-7 FPGA as programmable logic (PL) \cite{Zynq}, running on top of a Xilinx ZC706 evaluation board \cite{ZC706}; (ii) an Analog Devices (AD)-9361 RF transceiver \cite{AD9361} running on top of an AD-FMCOMMS2 evaluation board \cite{Fmcomms2}; (iii) four VERT2450 antennas \cite{Vert2450}, two for each TX/RX channel of the AD-9361; (iv) a Texas Instruments (TI) USB-TO-GPIO Interface Adapter to compute real-time power consumption of our board through the PMBUS standard \cite{USBToGPIO}. We chose this equipment since it provides significant flexibility in the both the RF, PL and PS components, and thus allows us to fully evaluate the trade-offs during system design.

\subsection{Deep Learning Model Training}\label{sec:res:mod_train}

As explained in Section \ref{sec:dlcore}, the first step in the \emph{RFLearn} system design process is to obtain a trained convolutional neural network (CNN) architecture. For this reason, we have performed a series of experiments with our testbed to obtain two datasets: (i) I/Q samples corresponding to 5 different modulation schemes (\textit{i.e.}, BPSK, QPSK, 8PSK, 16QAM, DQPSK); and (ii) I/Q samples of an OFDM transmission with three FFT size parameters (\textit{i.e.}, 64, 128, 256). To collect the samples, we have used another software-defined radio (\textit{i.e.}, a Xilinx Zedboard \cite{Zedboard} mounting an AD-FMCOMMS2 as RF transceiver) acting as transmitter, while our testbed was used to receive the samples.


If not stated otherwise, we train our model on inputs of size $32\times32\times2$ where $\ell = 32$, \textit{i.e.}, 32 rows of 32 I samples plus 32 rows of 32 Q samples. We train the model using Tensorflow for 20 epochs, using 150,000 samples per class. We use as test set an additional dataset of 200,000 inputs generated from the collected experimental data. The filter and pooling length has been set to 3, and the filter stride to 1.

\begin{figure}[!h]
    \centering
    \includegraphics[width=0.5\columnwidth,angle=-90]{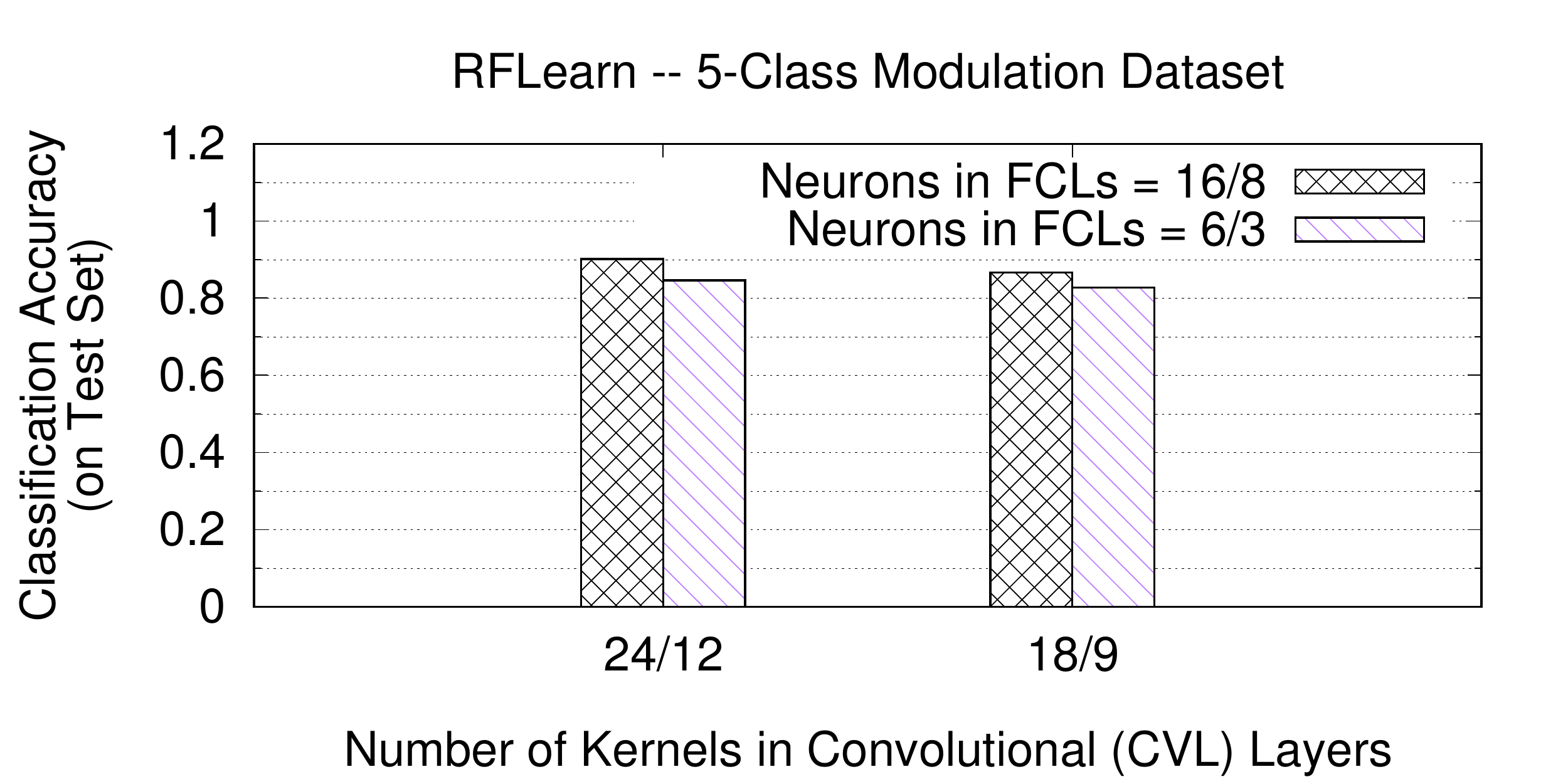}
    \caption{5-Class Modulation Dataset Accuracy Results.}
    \label{fig:mod_entire}
    \vspace{-0.3cm}
\end{figure}

To address \emph{ModRec}, we consider an architecture with $\mathrm{M = 2}$ and $\mathrm{K = 2}$, fixing $\mathrm{N = 1}$. Figure \ref{fig:mod_entire} shows the related classification accuracy. \textit{It can be observed that with a relatively small DL architecture with low number of kernels/neurons (as compared to modern computer vision models \cite{krizhevsky2012imagenet}) we can achieve an accuracy of at least 90\% over 5 classes.} This is also thanks to the shift-invariance property of CNNs. We can also conclude that the number of kernels and the number of neurons definitely impact the model's accuracy; by doubling the number of kernels and increasing the number of neurons from 6-3 to 16-8, we can increase the accuracy by about 14\%.


To further investigate the impact that the different kinds of modulations have on the model's accuracy, we trained the same DL architecture on two sets of 4 modulation classes, namely \textit{S1} = \{BPSK, QPSK, 16QAM, 8PSK\} and \emph{S2} = \{BPSK, QPSK, 16QAM, DQPSK\}. Since in \textit{S2} we consider two very similar modulations (\textit{i.e.}, QPSK and DQPSK), we should expect worse classification accuracy with respect to \textit{S1} with the same DL architecture.

\begin{figure}[!h]
    \centering
    \includegraphics[width=0.6\columnwidth,angle=-90]{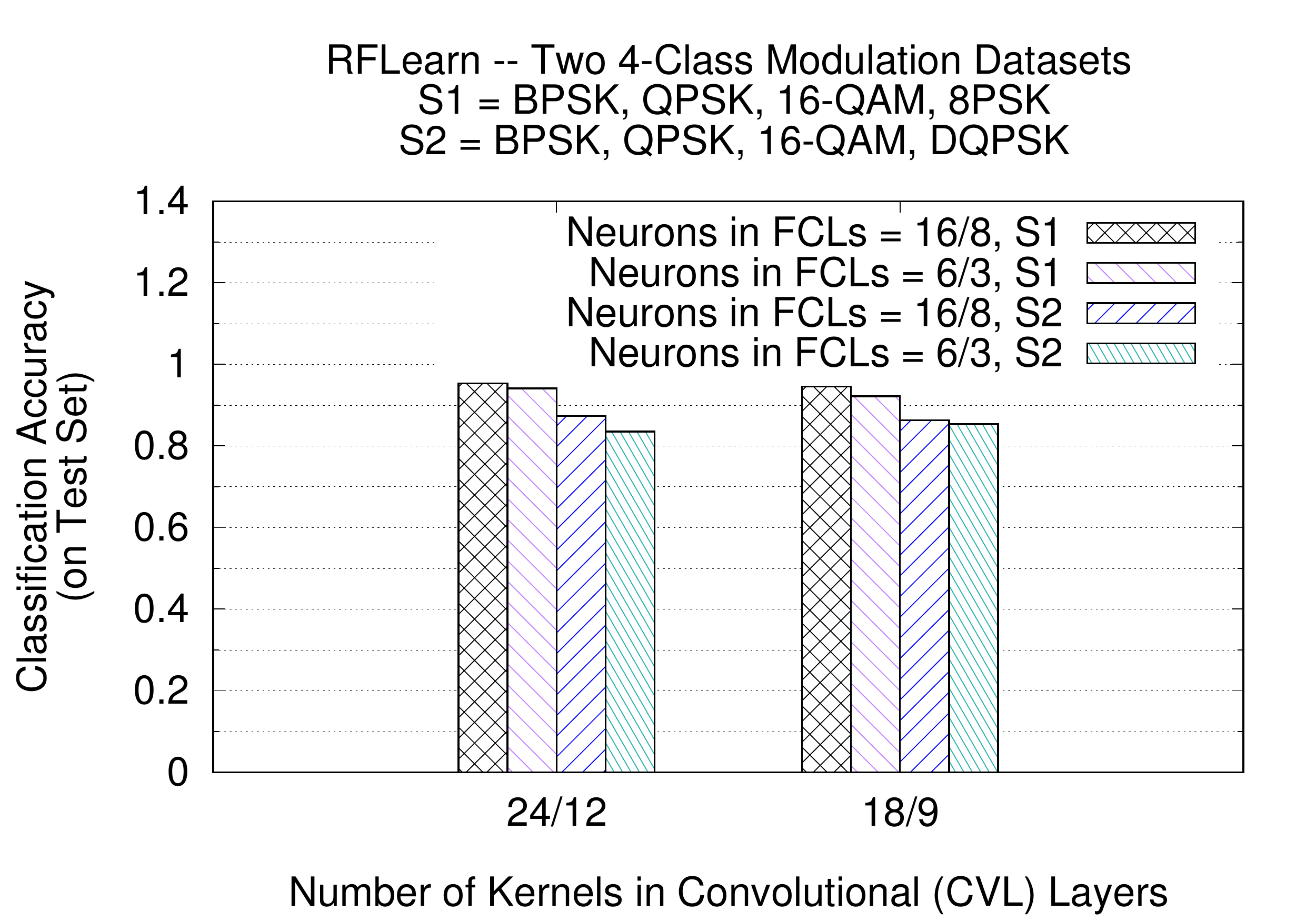}
    \caption{Two 4-Class Modulation Datasets Accuracy Results.}
    \label{fig:mod_subsets}
    \vspace{-0.3cm}
\end{figure}


Figure \ref{fig:mod_subsets} shows the model's accuracy for both \emph{S1} and \emph{S2}. As expected, Figure \ref{fig:mod_subsets} concludes that the model's accuracy is higher for \emph{S1} than for \emph{S2} (9\% on average), since the classes are more distinct in the former case. Therefore, not only does the number of modulation classes impact the model's accuracy, but also the \textit{type} of modulation classes considered.



\begin{figure}[!h]
    \centering
    \includegraphics[width=0.5\columnwidth,angle=-90]{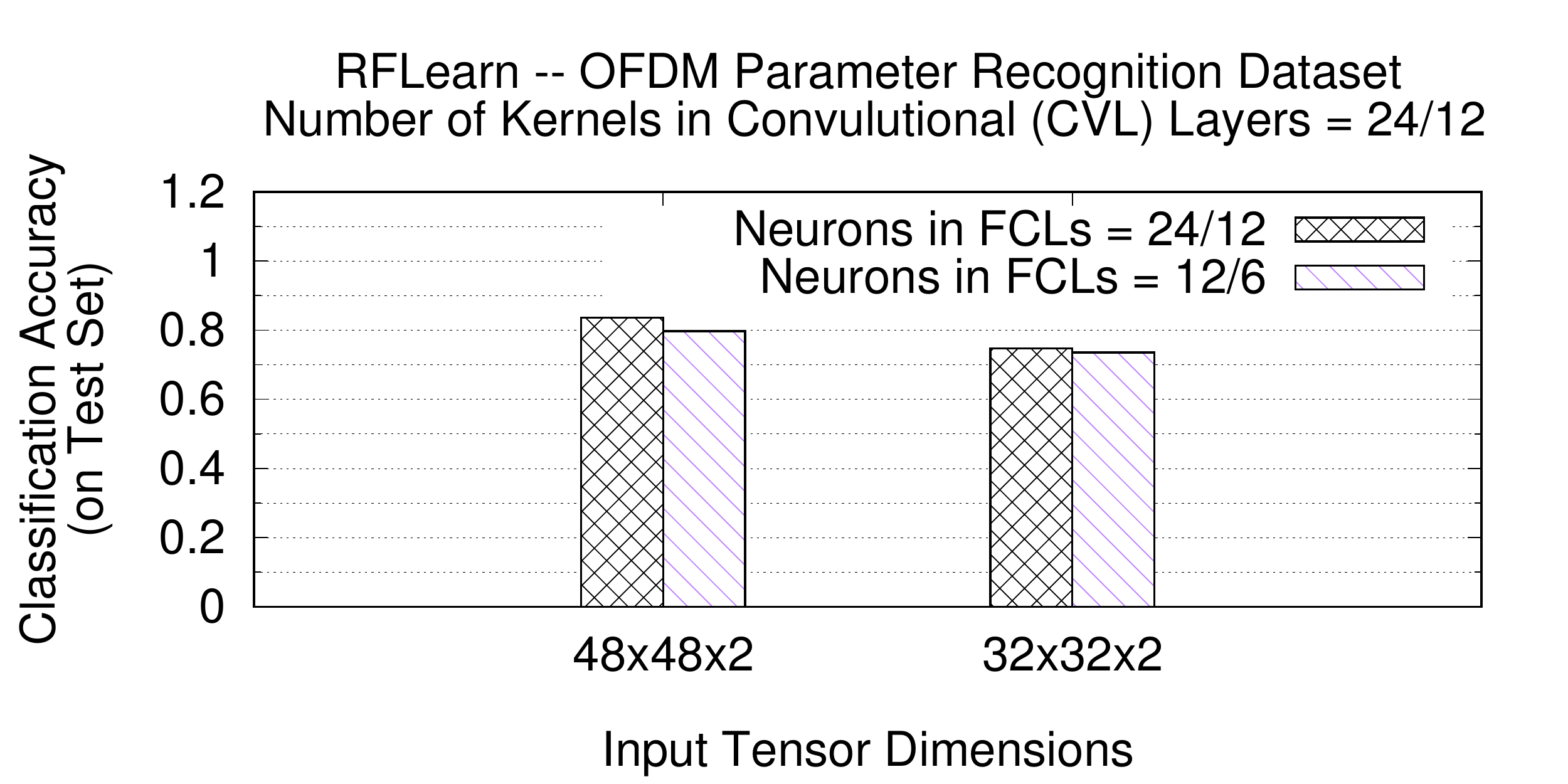}
    \caption{OFDM Recognition Dataset Accuracy Results.}
    \label{fig:ofdm}
    \vspace{-0.3cm}
\end{figure}

To investigate \emph{OFDMRec}, we have trained an architecture with a greater number of kernels/neurons and also increased the input size. Figure \ref{fig:ofdm} reports the classification results as a function of parameters and input size. \textit{As we can see, by increasing the input size to 48x48 we can increase the accuracy by 10\%, which concludes that an increase in model complexity increases classification accuracy accordingly.}

\subsection{RFLearn vs. Software Latency/Space/Power Comparison}\label{sec:res:hwvssw}

We now compare with our experimental testbed the \emph{RFLearn} latency performance vs. a software (SW) implementation. To this end, we have used the \emph{RFLearn HLS Library} to generate an equivalent model in C++ code to be executed in the PS portion of our testbed to test the difference in latency. To measure latency down to the clock cycle level, we have used an additional \textit{AXI Timer} core \cite{AXITimerCore} to count the number of clock cycles needed to produce the classification result in both hardware and software. To allow a fair comparison between the SW and the HW version, the testbed was run in ``baremetal'' mode (\textit{i.e.}, without operating system).
  
In the following experiments, we set the PL clock frequency to $\mathrm{100~MHz}$ (equivalent to $\mathrm{10ns}$ clock period), with the exception of the RF front-end core that is clocked at $\mathrm{200~MHz}$. Note that the frequency of each PL clock can be changed at any time through register configuration, without the need to implement each core in the PL from scratch. The CPU clock speed is instead $\mathrm{667~MHz}$ as per vendor datasheet. 

\renewcommand{\arraystretch}{1.2}
\begin{table}[h!]
\centering
\begin{tabular}{|c|c|c|c|c|c|c|}
\hline
\textbf{Kern} &  \textbf{Neur} & \textbf{SW} & \textbf{RFLearn} & \textbf{BRAM} & \textbf{LUT}   \\ \hline
\multirow{2}{*}{24} & 16 & 235.8ms & 13.7ms & 166 (15\%) & 28247 (12\%) \\ \cline{2-6}
                    & 8 & 220.1ms & 13.2ms & 166 (15\%) & 28227 (12\%) \\ \hline
\multirow{2}{*}{12} & 16 & 120.7ms & 6.9ms & 86 (7\%) & 20427 (9\%) \\ \cline{2-6}
                    & 8 & 111.9ms & 6.6ms & 86 (7\%) & 20406 (9\%) \\ \hline
\multirow{2}{*}{6} & 16 & 61.1ms & 3.4ms &  46 (4\%) & 16413 (7\%)\\ \cline{2-6}
                    & 8 & 56.5ms & 3.2ms & 46 (4\%) & 16399 (7\%)\\ \hline
\hline
\end{tabular}
\caption{RFLearn/SW Comparison, $\mathrm{M=1}$, $\mathrm{K=1}$.}
\label{tab:hw_sw_m1_k1}
\end{table}

Tables \ref{tab:hw_sw_m1_k1} and \ref{tab:hw_sw_m2_k2} report the RFLearn vs. SW comparison in terms of latency (expressed in milliseconds), and the related HW resource consumption (with related percentage) in terms of number of BRAM and look-up tables (LUT), for the $\mathrm{M=1}$, $\mathrm{K=1}$ and $\mathrm{M=2}$, $\mathrm{K=2}$ architectures, respectively. For the sake of space, we do not report the number of flip-flops (FF) consumed since it is about 1\% of the total resources in all the considered cases. The number of DSP48E1\footnote{A DSP48E1 is a complex circuit providing a multiplier, an accumulator, a pre-adder, and two arithmetic logic unit, among other features.} slices \cite{DSPSlice} consumed was 21 and 39 out of 900, respectively. For each SW latency measurement, we report the average over 100 repetitions. We do not report standard deviations since they are below 1\% of the average. 

\renewcommand{\arraystretch}{1.2}
\begin{table}[h!]
\centering
\begin{tabular}{|c|c|c|c|c|c|c|}
\hline
\textbf{Kern} &  \textbf{Neur} & \textbf{SW} & \textbf{RFLearn} & \textbf{BRAM} & \textbf{LUT}   \\ \hline
\multirow{2}{*}{24-12} & 16-8 & 1376.4ms & 75.9ms & 220 (20\%) & 23673 (10\%) \\ \cline{2-6}
                    & 6-3 & 1334.1ms & 75.6ms & 220 (20\%) & 23677 (10\%) \\ \hline
\multirow{2}{*}{18-9} & 16-8 & 767.8ms & 45.2ms & 220 (20\%) & 21738 (9\%) \\ \cline{2-6}
                    & 6-3 & 795.2ms & 44.9ms & 220 (20\%) & 21689 (9\%) \\ \hline
\multirow{2}{*}{12-6} & 16-8 & 389.17ms & 22.3ms &  116 (10\%) & 19636 (8\%)\\ \cline{2-6}
                    & 6-3 & 380.86ms & 22.1ms & 116 (10\%) & 19663 (8\%)\\ \hline
\hline
\end{tabular}
\caption{RFLearn/SW Comparison, $\mathrm{M=2}$, $\mathrm{K=2}$.}
\label{tab:hw_sw_m2_k2}
\end{table}

The first important result to remark is the significant different in latency performance between \emph{RFLearn} and SW. On the average, when $\mathrm{M=1}$, $\mathrm{K=1}$, \emph{RFLearn} improves the latency by about 17x, \textit{i.e.}, an order of magnitude with respect to SW, with a tolerable BRAM and LUT occupation of 15\% and 12\% in the worst case, respectively. The latency improvement brought by \emph{RFLearn} is confirmed also in the $\mathrm{M=2}$, $\mathrm{K=2}$ experiments , where the latency improvement with respect to SW continues to be about 17x on the average, at the cost of an increase in HW resource consumption (20\% vs 15\% BRAM in the worst case). Surprisingly enough, in some cases \emph{RFLearn} consumes less LUT resources when $\mathrm{M=2}$, $\mathrm{K=2}$. This can be explained by the fact that in these cases the \textit{Flatten} layer (used to transform a tensor input to a linear input to the FCL) has less inputs than with $\mathrm{M=1}$, $\mathrm{K=1}$, which causes less LUT consumption.

\renewcommand{\arraystretch}{1.2}
\begin{table}[h!]
\centering
\begin{tabular}{|c|c|c|c|c|c|c|}
\hline
\textbf{Exp} &  \textbf{1.0V} & \textbf{1.8V} & \textbf{1.5V} & \textbf{2.5V} & \textbf{3.3V} & \textbf{Total}  \\ \hline
Idle    & 0.16A & 0.06A & 0.02A & 0.11A  & 0.06V & 0.771W\\ \hline
Software & 0.28A & 0.12A & 0.03A & 0.11A & 0.06V & 1.014W \\ \hline  
\emph{RFLearn} & 0.37A & 0.13A & 0.03A &  0.13A & 0.06A & 1.172W \\  \hline  
\hline
\end{tabular}
\caption{RFLearn/SW/Idle Power Comparison.}
\label{tab:power_cons}
\end{table}

Table \ref{tab:power_cons} summarizes the current absorption (in Amperes) as measured at the different power rails of the ZC706 board. To obtain these results, we selected the 24-12-16-8 \emph{RFlearn} model (the most complex and thus, the worst case for power consumption) and averaged the results over 1000 measurements. As expected, \emph{RFLearn} experiences higher power consumption than the software-based implementation. However, the lower latency ($\mathrm{75.9~ms}$ vs $\mathrm{1376.4~ms}$) experienced by \emph{RFLearn} allows outstanding energy savings with respect to software. For example, in the considered case, the \emph{RFLearn} energy consumption is $\mathrm{87.9~mJ}$, which is about 15x lower than software ($\mathrm{1395.6~mJ}$).

\subsection{HLS Latency Optimization}\label{sec:res:opt}

We have mentioned in Section \ref{sec:dlcore} that \emph{RFLearn} is capable to decrease drastically the latency of the DL learning core through HLS optimization, at the cost of an increase in HW consumption. To prove this point, Table \ref{tab:opt_hls} shows the decrease in latency for different DL architectures upon HLS optimization, and the related amount of DSP48E1 slices consumed by the circuit. We do not report the increase in BRAM, LUT and FF since it was less than 1\% in all cases.

\renewcommand{\arraystretch}{1.2}
\begin{table}[h!]
\centering
\begin{tabular}{|c|c|c|c|}
\hline
\textbf{Kern} &  \textbf{Neur} & \textbf{Latency} & \textbf{DSP48E1}   \\ \hline
24 & 16  & 13.7ms $\rightarrow$ 8.2ms (-67\%) & \multirow{2}{*}{39 $\rightarrow$ 75 (+92\%)} \\ \cline{1-3}
3 & 16   & 1.6ms $\rightarrow$ 1.04ms (-54\%) &  \\ \hline
24-12 & 16-8  & 75.9ms $\rightarrow$ 37.9ms (-100\%) & \multirow{2}{*}{21 $\rightarrow$ 39 (+85\%)}\\ \cline{1-3}
12-6 & 16-8  & 22.3ms $\rightarrow$  11.5ms (-93\%) &  \\ \hline
\hline
\end{tabular}
\caption{\emph{RFLearn} Optimization, Latency vs. HW Space.}
\label{tab:opt_hls}
\vspace{-0.3cm}
\end{table}

The optimization made through HLS was to pipeline the loops corresponding to the computation of one filter output, so that the summing operations in Equation \ref{eq:filters} can be executed in parallel. Table \ref{tab:opt_hls} shows that by pipelining the convolution loops, we can achieve a significant reduction in latency. We point out that the decrease in latency becomes ever more evident as (i) the number of convolutional layers (CVLs) and (ii) the number of kernels in one layer increase. Indeed, we have a 67\% vs. 100\% latency reduction when going from one to two CVLs, and a 67\% vs. 54\% by going from 24 to 3 kernels. Obviously, this decrease in latency corresponds to an increase in DSP48E1 circuitry, which is almost double in the first architecture. Although the SoC considered in this paper supports up to 900 DSP48E1s, other architectures might have less DSP circuitry. Therefore, the trade-off between space and latency must always be considered before deploying the architecture on the SoC.




\section{Related Work}\label{sec:rw}

The usage of supervised machine learning  techniques to interpret the wireless spectrum has been extensively investigated over the last few years; the reader can refer to \cite{Bkassiny-ieeecommsurtut2013,Jiang-ieeewcomm2017,Chen-arxiv2017} for excellent surveys on the topic.  Most of existing work is based on traditional low-dimensional machine learning \cite{Wong-isspa2001,Xu-ieeetvt2010,Pawar-ieeetifs2011,Shi-ieeetcomm2012,Ghodeshar-icscn2015}, which requires (i) extraction and careful selection of complex features from the RF waveform (\textit{i.e.}, average, median, kurtosis, skewness, high-order cyclic moments, etc.); and (ii) the establishment of tight decision bounds between classes based on the current application, which are derived either from mathematical analysis or by learning a carefully crafted dataset \cite{shalev2014understanding}. In other words, since feature-based machine learning is (a) significantly application-specific in nature; and (b) it introduces additional latency and computational burden due to feature extraction, its application to real-time hardware-based wireless spectrum analysis becomes unpractical, as the wireless radio hardware should be changed according to the specific application under consideration.

Recent advances in deep learning \cite{lecun2015deep} have prompted researchers to investigate whether similar techniques can be used to analyze the sheer complexity of the wireless spectrum. For a compendium of existing research on the topic, the reader can refer to \cite{Mao-ieeecomm2018}. Among other advantages, deep learning is significantly amenable to be used for real-time hardware-based spectrum analysis, since different model architectures can be reused to different problems as long as weights and hyper-parameters can be changed through software. Among other issues, physical-layer modulation recognition through deep learning has received significant attention in the last two years \cite{OShea-ieeejstsp2018,o2017introduction,wang2017deep,West-dyspan2017,Kulin-ieeeaccess2018,Karra-ieeedyspan2017}.  O'Shea \emph{et al.} present in \cite{OShea-ieeejstsp2018} several deep learning models to address the modulation recognition problem, while in \cite{Karra-ieeedyspan2017} Karra \emph{et al.} train hierarchical deep neural networks to identify data type, modulation class and modulation order. Kulin \emph{et al.} present in \cite{Kulin-ieeeaccess2018} a conceptual framework for end-to-end wireless deep learning, followed by a comprehensive overview of the methodology for collecting spectrum data, designing wireless signal representations, forming training data
and training deep neural networks for wireless signal classification tasks.

The core issue with prior approaches is that they leverage deep learning to perform offline spectrum analysis only. On the other hand, the opportunity of real-time hardware-based spectrum knowledge inference remains substantially uninvestigated. For this reason, this paper proposes a hardware architecture and learning core design strategy that together bring the power of deep learning \emph{directly to the RF hardware loop}, which will enable sophisticated, real-time decision-making and knowledge inference with limited human intervention.

\section{Conclusions}\label{sec:rw}

In this paper, we have proposed \emph{RFLearn}, the first learning-in-the-RF-loop system that enables spectrum-driven decisions through real-time hardware-based I/Q deep learning algorithms. We have provided a complete hardware architecture for \emph{RFLearn}, as well as a novel framework for RF deep learning circuit design that translates and optimizes the software-based implementation to produce a \emph{RFLearn}-compliant circuit. We have extensively evaluated  \emph{RFLearn} on a practical testbed by considering the problem of modulation and OFDM parameter recognition through deep learning, and explored in details the trade-off between hardware space vs. latency and model complexity vs. accuracy. Furthermore, we have compared the latency performance of \emph{RFLearn} with respect to a software-based system. Experimental results have demonstrated that \emph{RFlearn} decreases the latency and power consumption by respectively 17x and 15x with a relatively low hardware resource consumption.




\section*{Acknowledgements}

We sincerely thank the anonymous reviewers for their precious feedback, which helped us improve significantly the quality of our final manuscript. This work was supported in part by the National Science Foundation (NSF) under Grant CNS-1618727. The views and conclusions contained in this document are those of the authors and should not be interpreted as representing the official policies, either expressed or implied, of the National Science Foundation or the U.S.~Government.

\footnotesize

\end{document}